\documentclass[twocolumn]{revtex4-1}
\usepackage{amsmath,amssymb,amsfonts}
\usepackage{mathtools}
\usepackage{multirow}
\usepackage{graphicx}
\usepackage{dcolumn}
\usepackage{bm}
\usepackage{footmisc}
\usepackage[english]{babel}
\usepackage{natbib}
\usepackage{hyperref}
\usepackage[normalem]{ulem}
\usepackage{comment}
\usepackage{xcolor}
\usepackage{caption}
\usepackage{booktabs}

\preprint{APS/123-QED}
\begin{document}
\renewcommand{\arraystretch}{1.75}

\title{Rapid Climate Model Downscaling to Assess\\Risk of Extreme Rainfall in Bangladesh in a Warming Climate }

\author{\begin{tabular}{ccc}
Anamitra Saha && Sai Ravela\\
anamitra@mit.edu && ravela@mit.edu\\
&Earth Signals and Systems Group&\\
\end{tabular}\\
Earth, Atmospheric and Planetary Sciences \\ 
Massachusetts Institute of Technology\\
Cambridge, MA, USA 
}

\thispagestyle{plain}
\pagestyle{plain}

\sloppy
\begingroup
\hyphenpenalty 10000

\begin{abstract}
As climate change drives an increase in global extremes, it is critical for Bangladesh, a nation highly vulnerable to these impacts, to assess future risks for effective adaptation and mitigation planning. Downscaling coarse-resolution climate models to finer scales is essential for accurately evaluating the risk of extremes. In this study, we apply our downscaling method, which integrates data, physics, and machine learning, to quantify the risks of extreme precipitation in Bangladesh. The proposed approach successfully captures the observed spatial patterns and risks of extreme rainfall in the current climate while generating risk and uncertainty estimates by efficiently downscaling multiple models under future climate scenarios. Our findings project that the risk of extreme rainfall rises across the country, with the most significant increases in the northeastern hilly and southeastern coastal areas. Projections show that the daily maximum rainfall for a 100-year return period could increase by approximately 50 mm/day by mid-century and around 100 mm/day by the end of the century. However, substantial uncertainties remain due to variations across multiple climate models and scenarios.
\end{abstract}

\maketitle

\section{Introduction}
Bangladesh, a densely populated developing nation, is highly vulnerable to climate extremes. Over the past several decades, it has faced significant economic losses and human casualties from extreme events such as tropical cyclones, floods, and heat waves. With the global increase in extreme weather events driven by climate change, policymakers in Bangladesh need to understand how these risks may evolve in the coming decades to inform adaptation and mitigation strategies. Accurately assessing future risks and associated uncertainties is crucial for ensuring food security, water availability, and the protection of public infrastructure in Bangladesh.

State-of-the-art climate models, such as those from Coupled Model Intercomparison Project Phase 6 (CMIP6), are too coarse to represent the finer-scale geophysical processes essential to quantifying risk. Coupled with model biases, they typically underestimate extremes and, thus, risks. Running these models in high resolution is computationally expensive, which has driven substantial interest in downscaling coarse model outputs to achieve finer resolutions.

Downscaling methods typically fall into two broad approaches: theory-driven (physics-based) and data-driven strategies. Both approaches have inherent limitations. Physics-based numerical techniques (i.e., dynamical downscaling) are computationally intensive, and the often employed process parameterizations also produce bias. In contrast, data-driven approaches, such as statistical and machine learning techniques, are rapid and efficient but are usually limited by data scarcity and may not adhere to geophysical principles. In Saha and Ravela (2024a), we introduced a rapid large-ensemble downscaling framework that integrates statistics, simplified physics, and adversarial learning to address some of these limitations~\cite{saha2024statistical}. This approach performs better than purely physics-based or data-driven models while being computationally efficient.

In a subsequent study, we applied this methodology to assess future extreme precipitation risk in Bangladesh, focusing on the HighResMIP experiment, which is limited to a single climate-socioeconomic scenario (SSP5-8.5) up to the mid-21st century~\cite{saha2024rapid}. However, it is crucial to account for uncertainties across multiple scenarios and capture model errors for effective policy planning. In this study, we extend the previous work by downscaling outputs from thirteen climate models within the Scenario-MIP experiment of CMIP6, covering four scenarios through the end of the 21st century. This provides a comprehensive assessment of Bangladesh's projected changes in extreme precipitation risk and associated uncertainties across models and scenarios. The findings suggest a nationwide increase in risk, particularly in the northeastern hilly region and southeastern coastal areas. However, significant uncertainties remain in model projections, highlighting the need to communicate these uncertainties for better-informed adaptation and mitigation strategies in Bangladesh.

The rest of the manuscript is structured as follows: Section~\ref{section:methods} briefly outlines the data and methodology, while Section~\ref{section:results} presents the results. Finally, Section~\ref{section:discussion} discusses the study’s implications, limitations, and future directions. For a more comprehensive description of the methodology and prior research, refer to Saha and Ravela (2024a, 2024b)~\cite{saha2024statistical, saha2024rapid}.

\begin{table}[htpb]
    \centering
    \captionof{table}{List of CMIP6 models used in this study}
    \begin{tabular}{|l|c|}
    \hline
        \multicolumn{1}{|c|}{\textbf{Models}} & \textbf{Resolution} \\[1.1ex]
    \hline\hline
        BCC-CSM2-MR & 112.5 Km \\
    \hline 
        CMCC-ESM2 & 100$\times$125 Km \\
    \hline
        GFDL-ESM4 & 100$\times$125 Km \\
    \hline
        KACE-1-0-G & 125$\times$187.5 Km \\
    \hline
        MPI-ESM1-2-HR & 100 Km \\
    \hline
        MRI-ESM2-0 & 112.5 Km \\
    \hline
        ACCESS-ESM1-5 & 125$\times$187.5 Km \\
    \hline
        CMCC-CM2-SR5 & 100$\times$125 Km \\
    \hline
        EC-Earth3 & 70 Km \\
    \hline
        IITM-ESM & 187.5 Km \\
    \hline
        IPSL-CM6A-LR & 125$\times$250 Km \\
    \hline
        MIROC6 & 140 Km \\
    \hline
        MPI-ESM1-2-LR & 187.5 Km \\
    \hline
    \end{tabular}
    \label{tab:scenariomip}
\end{table}

\begin{table}[htpb]
    \centering
    \captionof{table}{List of Shared Socioeconomic Pathways used in this study}
    \begin{tabular}{|l|p{0.1\textwidth}|p{0.1\textwidth}|p{0.1\textwidth}|}
    \hline
        \multicolumn{1}{|c|}{\textbf{SSP}} & \textbf{Forcing} & \textbf{Adaptation \newline Challenge} & \textbf{Mitigation \newline Challenge} \\[1.1ex]
    \hline\hline
        SSP1-2.6 & Low & Low & Low \\
    \hline 
        SSP2-4.5 & Medium & Medium & Medium \\
    \hline
        SSP3-7.0 & High & High & High \\
    \hline
        SSP5-8.5 & High & Low & High \\
    \hline
    \end{tabular}
    \label{tab:ssp}
\end{table}

\begin{figure}[htpb]
\centering
\includegraphics[width=7cm]{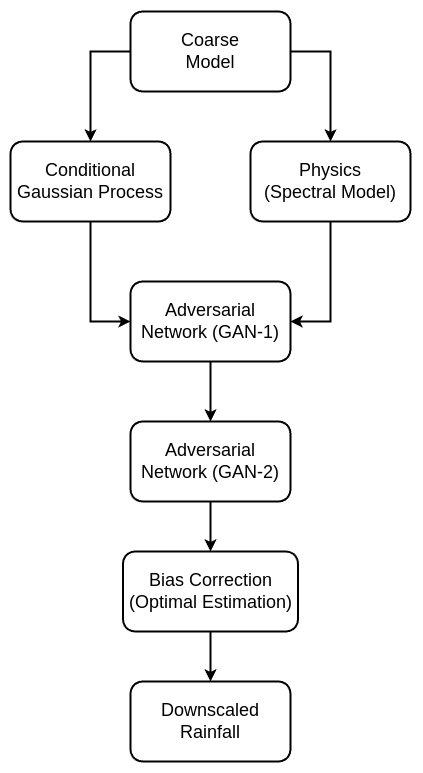}
\caption{A schematic representation of the downscaling method. Coarse-resolution model predictions are downscaled to the resolution of high-resolution observation with the help of a simple statistical model (conditional Gaussian process), simplified physics (spectral model), and two-step adversarial networks (GAN-1 and GAN-2). Optimal estimation bias corrects the downscaled rainfall in the present climate. (From Saha and Ravela, 2024b, Figure 1~\cite{saha2024rapid})}
\label{fig:method}
\end{figure}

\begin{figure*}[htpb]
\centering
\includegraphics[width=13cm]{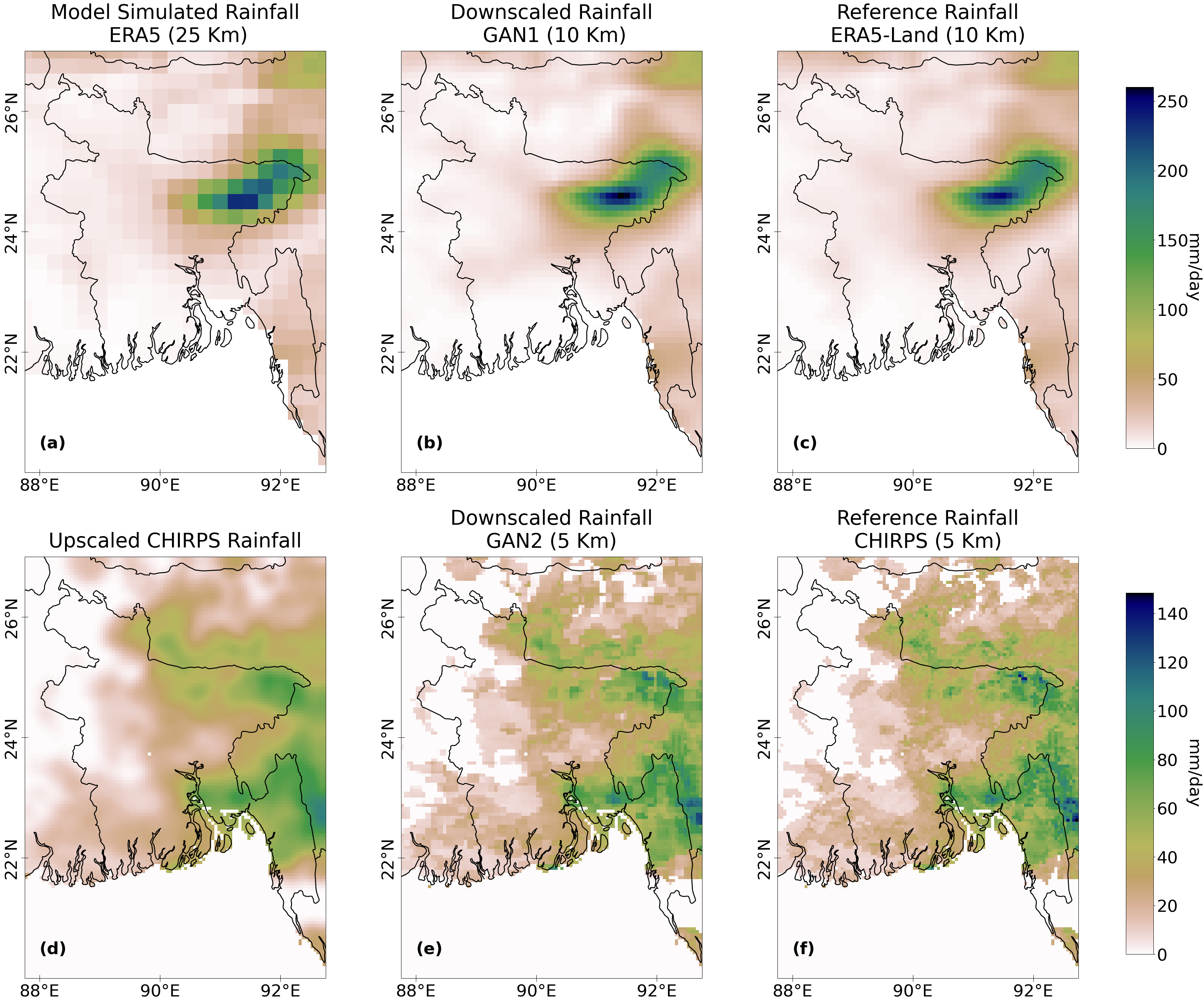}
\caption{A qualitative comparison of coarse, downscaled, and fine-resolution reference rainfall fields. The top row features an extreme event from ERA5/ERA5-Land and corresponding downscaled rainfall from GAN-1. The bottom row includes an extreme event from CHIRPS and corresponding downscaled rainfall from GAN-2. The panels are as follows: (a) Low-resolution ($0.25^{\circ}$) rainfall from ERA5, (b) downscaled rainfall from GAN-1 at resolution $0.1^{\circ}$,\newline (c) reference high-resolution ($0.1^{\circ}$) rainfall from ERA5-Land, (e) rainfall upscaled from CHIRPS,\newline (f) downscaled rainfall from GAN-2 at resolution $0.05^{\circ}$, (g) original high-resolution ($0.05^{\circ}$) reference rainfall from CHIRPS. A comparison of the middle and right columns highlights the effectiveness of our downscaling model. (From Saha and Ravela, 2024b, Figure 2~\cite{saha2024rapid})}
\label{fig:downscalingmodeleval}
\end{figure*}

\begin{figure}[htpb]
\centering
\includegraphics[width=8cm]{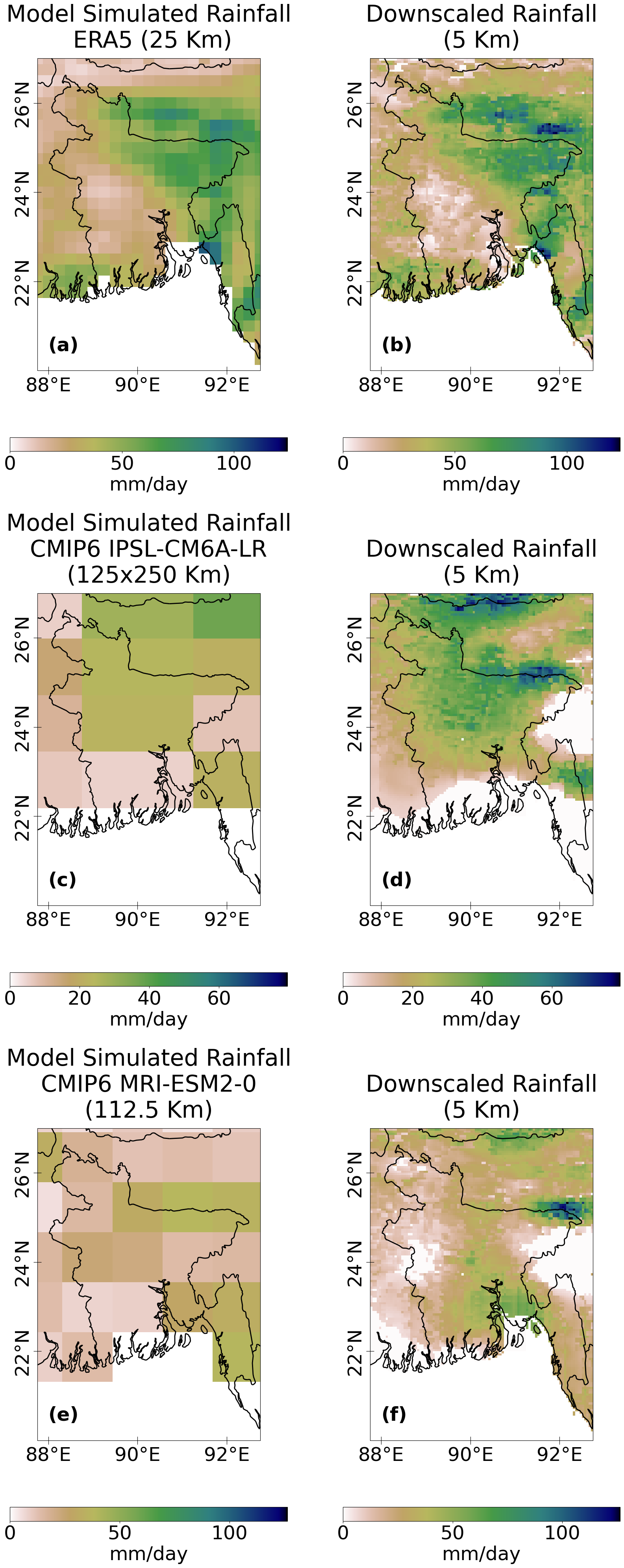}
\caption{The left column shows the coarse rainfall fields simulated by the models, while the right column shows the corresponding downscaled rainfall fields. (a) Coarse rainfall fields simulated by ERA5 Reanalysis model (resolution $0.25^{\circ}$), (b) corresponding downscaled rainfall field. (c-d) CMIP6 model IPSL-CM6A-LR (resolution $1.25^{\circ}\times2.5^{\circ}$), (e-f) CMIP6 model MRI-ESM2-0 (resolution $1.125^{\circ}$). Although a $0.25^{\circ}$ resolution model trains downscaling, it can effectively downscale coarser climate models.}
\label{fig:downscalingapplication}
\end{figure}

\begin{figure*}[htpb]
\centering
\includegraphics[width=15cm]{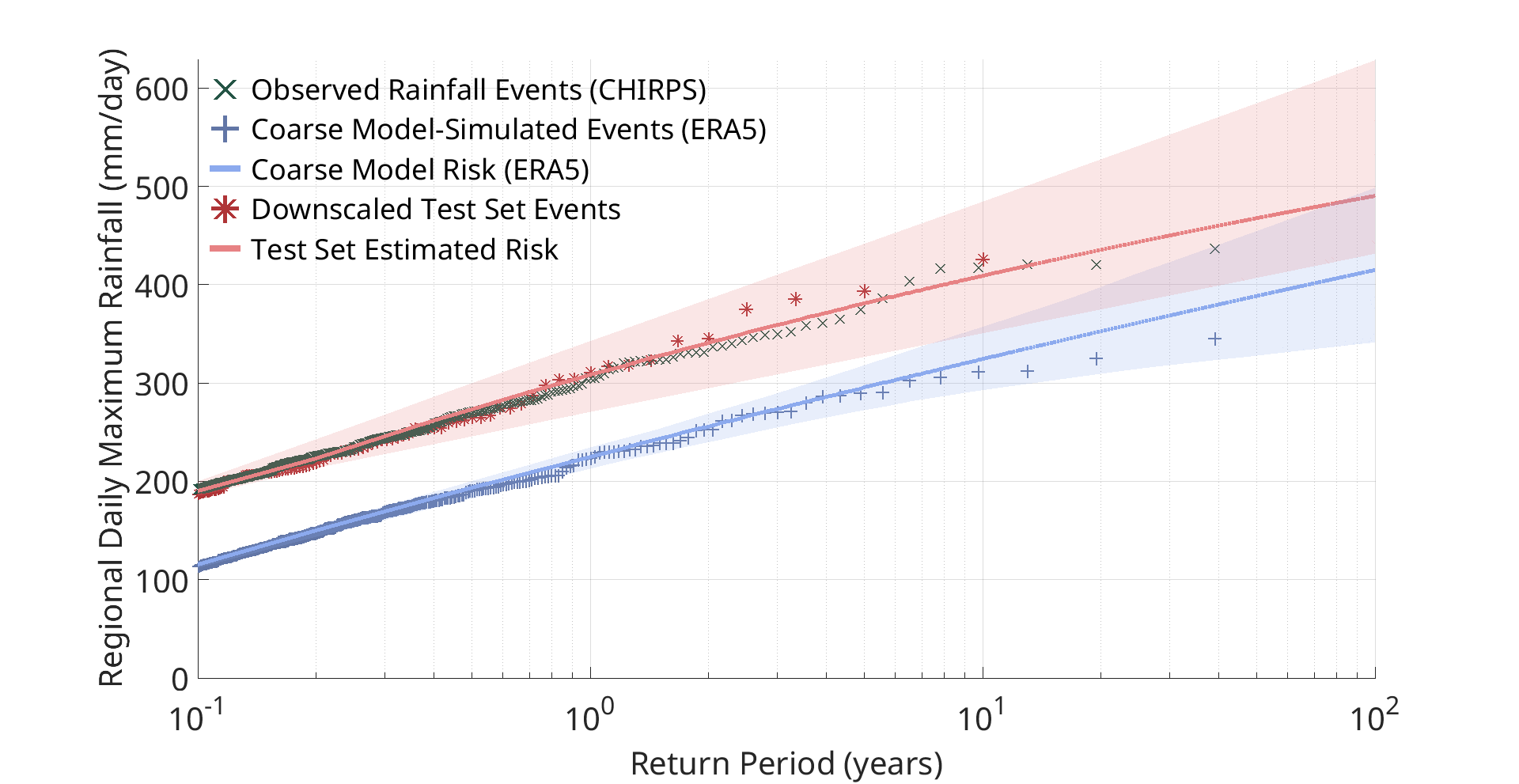}
\caption{Comparison of extreme rainfall risks as captured by a coarse model (ERA5), high-resolution observations (CHIRPS), and downscaled rainfall data in the present climate. The figure displays daily regional rainfall maxima plotted against their return periods, with individual events from ERA5 shown as blue plus signs, CHIRPS as green crosses, and downscaled rainfall as red asterisks. Solid lines represent Generalized Pareto fits through these events, extended to a 100-year return period, with the shaded area indicating uncertainty. This figure illustrates that coarse models tend to underestimate extreme rainfall risks, but downscaling can better capture the observed risk in the present climate. (From Saha and Ravela, 2024b, Figure 4~\cite{saha2024rapid})}
\label{fig:riskpresent}
\end{figure*}

\begin{figure}[htpb]
\centering
\includegraphics[width=9cm]{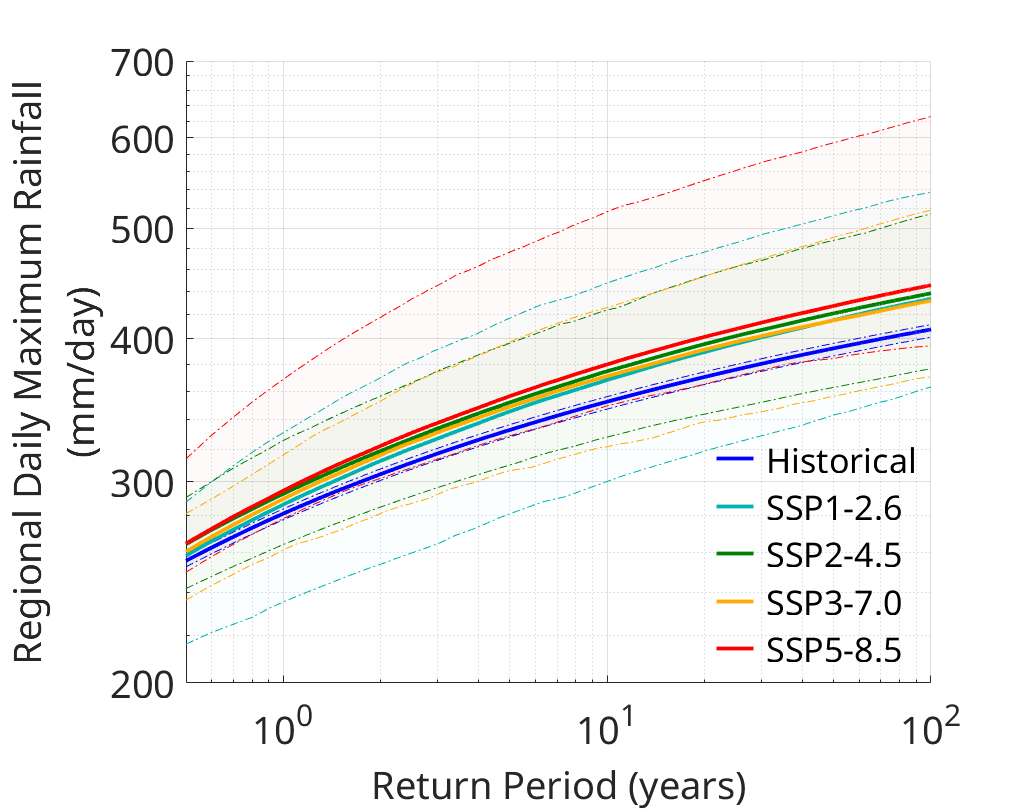}
\caption{Comparison of extreme rainfall risks captured by downscaled CMIP6 models for the present climate (1985-2014, shown in blue) and four future climate scenarios at the mid-century (2031-2050): SSP1-2.6 (cyan), SSP2-4.5 (green), SSP3-7.0 (orange), and SSP5-8.5 (red). Solid lines denote the mean of Generalized Pareto fits for daily regional rainfall maxima from individual models, plotted against their return periods. The shaded area represents the inter-model spread. This figure displays the expected increase in the extreme rainfall risk from the present to the future climate and the model and scenario uncertainty around it.}
\label{fig:riskfuturemid}
\end{figure}

\begin{figure}[htpb]
\centering
\includegraphics[width=9cm]{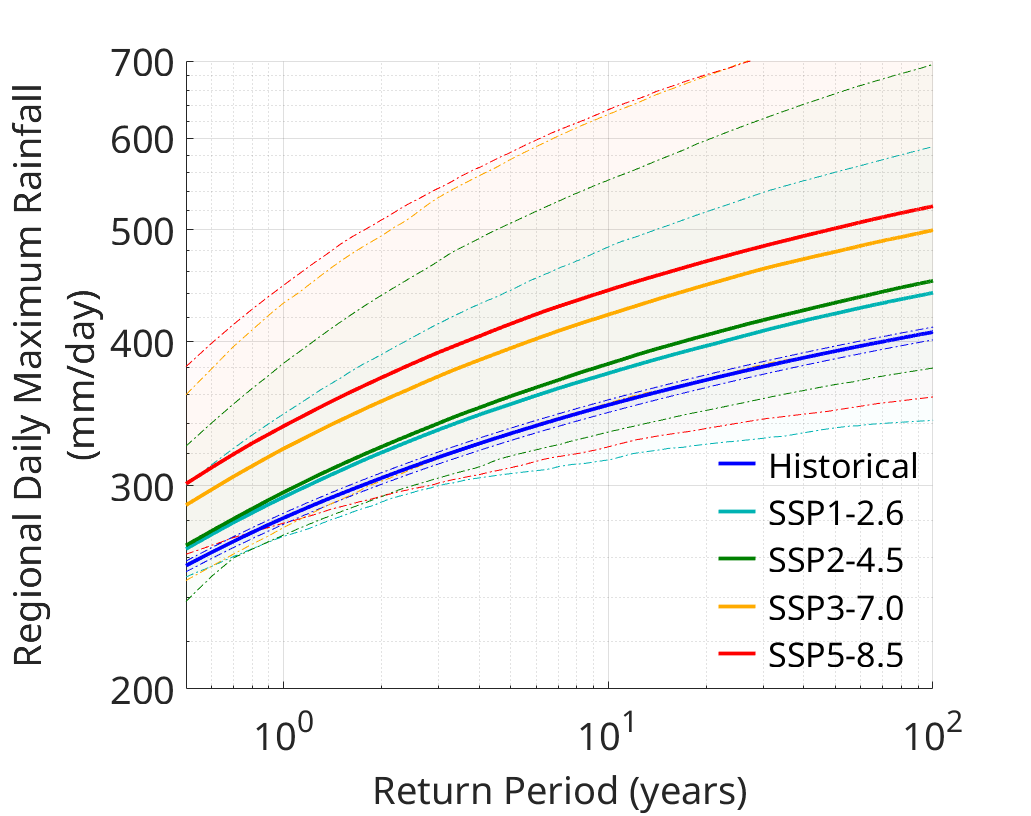}
\caption{Comparison of extreme rainfall risks captured by downscaled CMIP6 models for the present climate (1985-2014, shown in blue) and four future climate scenarios at the end of the century (2081-2100): SSP1-2.6 (cyan), SSP2-4.5 (green), SSP3-7.0 (orange), and SSP5-8.5 (red). Solid lines denote the mean of Generalized Pareto fits for daily regional rainfall maxima from individual models, plotted against their return periods. The shaded area represents the inter-model spread. This figure displays the expected increase in the extreme rainfall risk from the present to the future climate and the model and scenario uncertainty around it.}
\label{fig:riskfutureend}
\end{figure}

\section{Data and Methods}
\label{section:methods}
We downscale low-resolution ($0.25^{\circ}$ to $1^{\circ}$) model-simulated data to high-resolution ($0.05^{\circ}$) rainfall fields that are comparable to the gridded daily rainfall fields from Climate Hazards Group InfraRed Precipitation with Station data (CHIRPS). CHIRPS is a quasi-global rainfall dataset incorporating rainfall estimates from rain gauges and satellite observations~\cite{funk2015climate}. Our downscaling model trains using low-resolution data from the European Centre for Medium-Range Weather Forecasts (ECMWF) Reanalysis (ERA5), which is available at $0.25^{\circ}$ resolution~\cite{hersbach2020era5}. 

Due to biases in ERA5 and CHIRPS, their rainfall fields do not align daily, presenting a challenge in training a downscaling function. A two-stage downscaling process addresses this. Following the downscaling approach developed in SR24~\cite{saha2024statistical}, in the first stage (GAN-1), ERA5 data ($0.25^{\circ}$) downscales to ERA5-Land resolution ($0.1^{\circ}$). ERA5-Land is a high-resolution replay of the land surface component of ERA5, providing finer spatial detail~\cite{munoz2021era5}. In the second stage (GAN-2), upscaled CHIRPS rainfall fields become predictors for the original CHIRPS rainfall fields ($0.05^{\circ}$) as predictands. Figure~\ref{fig:method} presents a schematic representation of our downscaling approach. The approach consists of the following steps:
\begin{enumerate}
    \item A fast statistical model (iterative ensemble-approximated conditional Gaussian process regressor) to derive a ``first-guess" downscaled rainfall field~\cite{ravela2007fast,saha2024statistical,trautner2020informative}.
    \item A simplified physics-based model (spectral method) to estimate orography-induced precipitation~\cite{smith2004linear}.
    \item an adversarial network (GAN-1) that combines these two fields and produces an improved downscaled rainfall field at $0.1^{\circ}$ resolution~\cite{saha2024statistical}.
    \item Another adversarial network (GAN-2) that downscales the rainfall field to $0.05^{\circ}$ resolution~\cite{saha2024rapid}.
    \item An optimal estimation-based bias correction~\cite{saha2024rapid}.
\end{enumerate}

Priming the downscaling model with statistics and physics-derived rainfall fields~\cite{smith2004linear} improves the physical consistency and alleviates data paucity issues. Two-step adversarial learning addresses the lack of correspondence between model and data and captures high-frequency details in the downscaled field. Biases between the model and data are corrected with optimal estimation~\cite{rodgers1976retrieval} to ensure that the downscaled rainfall captures the observed risk of extremes in the present climate.

All the aforementioned datasets were obtained for the years 1981 to 2019 and split into three groups: training (1981-1999), validation (2000-2009), and testing (2010-2019). Additionally, simulated outputs of thirteen models from the Scenario Model Intercomparison Project (ScenarioMIP)~\cite{o2016scenario} from CMIP6 are obtained for one historical scenario (1951-2014) and four tier-1 future scenarios (2015-2100), namely SSP5-8.5, SSSP3-7.0, SSP2-4.5, and SSP1-2.6. Table~\ref{tab:scenariomip} lists the ScenarioMIP CMIP6 models used, and Table~\ref{tab:ssp} lists additional details on the SSP scenarios. The CMIP6 models are coarser than ERA5; a bicubic interpolation brings them to the exact resolution as ERA5 before applying the conditional Gaussian process. The same downscaling function is used for present and future climate projections, assuming that the downscaling function remains unchanged in the warming scenario. To assess the changes between historical and future periods, we define 1985-2014 as the present climate, 2031-2050 as the future climate mid-century, and 2081-2100 as the future climate end-century.

\section{Results}
\label{section:results}

Figure~\ref{fig:downscalingmodeleval} provides a qualitative comparison of our downscaling model's performance against reference rainfall fields ("truth"). Specifically, Figure~\ref{fig:downscalingmodeleval}a shows an extreme rainfall event from the ERA5 dataset at $0.25^{\circ}$ resolution, which GAN-1 subsequently downscales to $0.1^{\circ}$, see Figure~\ref{fig:downscalingmodeleval}b, compared with the corresponding rainfall field produced by ERA5-Land ($0.1^{\circ}$) in Figure~\ref{fig:downscalingmodeleval}c. Similarly, a rainfall event from the CHIRPS dataset ($0.05^{\circ}$), shown in Figure~\ref{fig:downscalingmodeleval}f, is upscaled in Figure~\ref{fig:downscalingmodeleval}d, and then downscaled to Figure~\ref{fig:downscalingmodeleval}e by GAN-2. Comparing the outcomes of both GAN-1 and GAN-2 with their respective references demonstrates that our model sufficiently captures the spatial pattern and heterogeneity of the high-resolution rainfall.

Figure~\ref{fig:downscalingapplication} showcases downscaling applied to different coarse-resolution climate model outputs of varying resolutions. A direct comparison with a reference rainfall field is impossible here due to a lack of correspondence between the model and observation on a daily scale. However, we observe that the downscaled rainfall exhibits a strong resemblance to high-frequency CHIRPS rainfall data, indicating the effectiveness of our approach. The downscaling method remains robust, even when applied to climate models at resolutions coarser than the $0.25^{\circ}$ data used for training.

Figure~\ref{fig:riskpresent} assesses the ability of downscaled rainfall data to reflect observed risk in the present climate. This figure presents the daily maxima of extreme events from CHIRPS, ERA5, and downscaled ERA5 rainfall against their return periods. Solid lines show the two-parameter Generalized Pareto distributions fit through them, with uncertainty around these lines estimated through bootstrapping. Notably, the coarse-resolution ERA5 model underestimates the risk of extremes, highlighting the necessity for downscaling. The results demonstrate that our downscaling approach reliably captures the observed risk. Given the model's effectiveness in representing both spatial patterns and extreme event risks in the current climate, we extend its application to both present and future climate projections from CMIP6 models under the assumption that the downscaling function remains invariant to changes in climate.

Figure~\ref{fig:riskfuturemid} compares the extreme rainfall risks captured by downscaled rainfall from thirteen models for the present and four future climate scenarios by the mid-century (2031-2050). Similarly, Figure~\ref{fig:riskfutureend} presents the risk by the end of the century (2081-2100). Since present climate data is individually bias-corrected against observations for each model, the variation between models is minimal. However, there is less consensus among the models in the future climate, as indicated by a more extensive inter-model spread. Most models project an increase in extreme risk for the future climate. The expected rise in extreme is the lowest in the sustainability scenario (SSP1-2.6) and highest in the fossil fuel-driven development scenario (SSP5-8.5). For the SSP5-8.5 scenario, approximately 50 mm/day in the return level of a 100-year return period is expected by the mid-century and 100 mm/day by the end of the century. Note that more sustainable scenarios like SSP1-2.6 and SSP2-4.5 show little to no increase in risk from mid-century to end-century, emphasizing the importance of limiting warming to these levels.

\begin{figure*}[t]
\centering
\includegraphics[width=16cm]{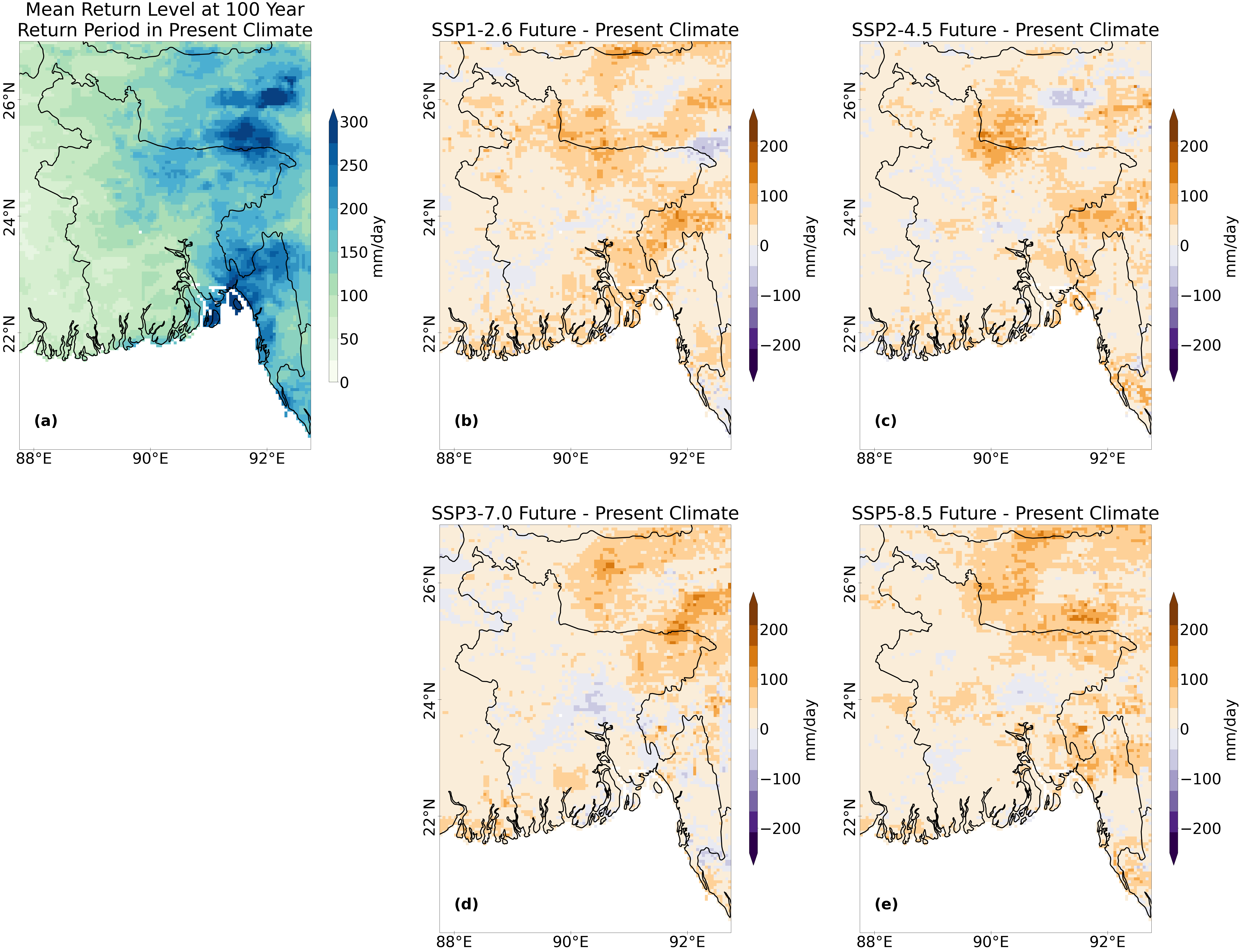}
\caption{Spatial distribution of extreme rainfall risk by the mid-century, captured by downscaled climate models. Panel (a) shows the mean return level at a 100-year return period for the present climate (1985-2014). Panel (b-e) displays the differences in the mean return level at a 100-year return period between the future (2031-2050) and present climates for four different scenarios: (b) SSP1-2.6, (c) SSP2-4.5, (d) SSP3-7.0, and (e)SSP5-8.5.}
\label{fig:riskfuturespatialmid}
\end{figure*}

\begin{figure*}[t]
\centering
\includegraphics[width=16cm]{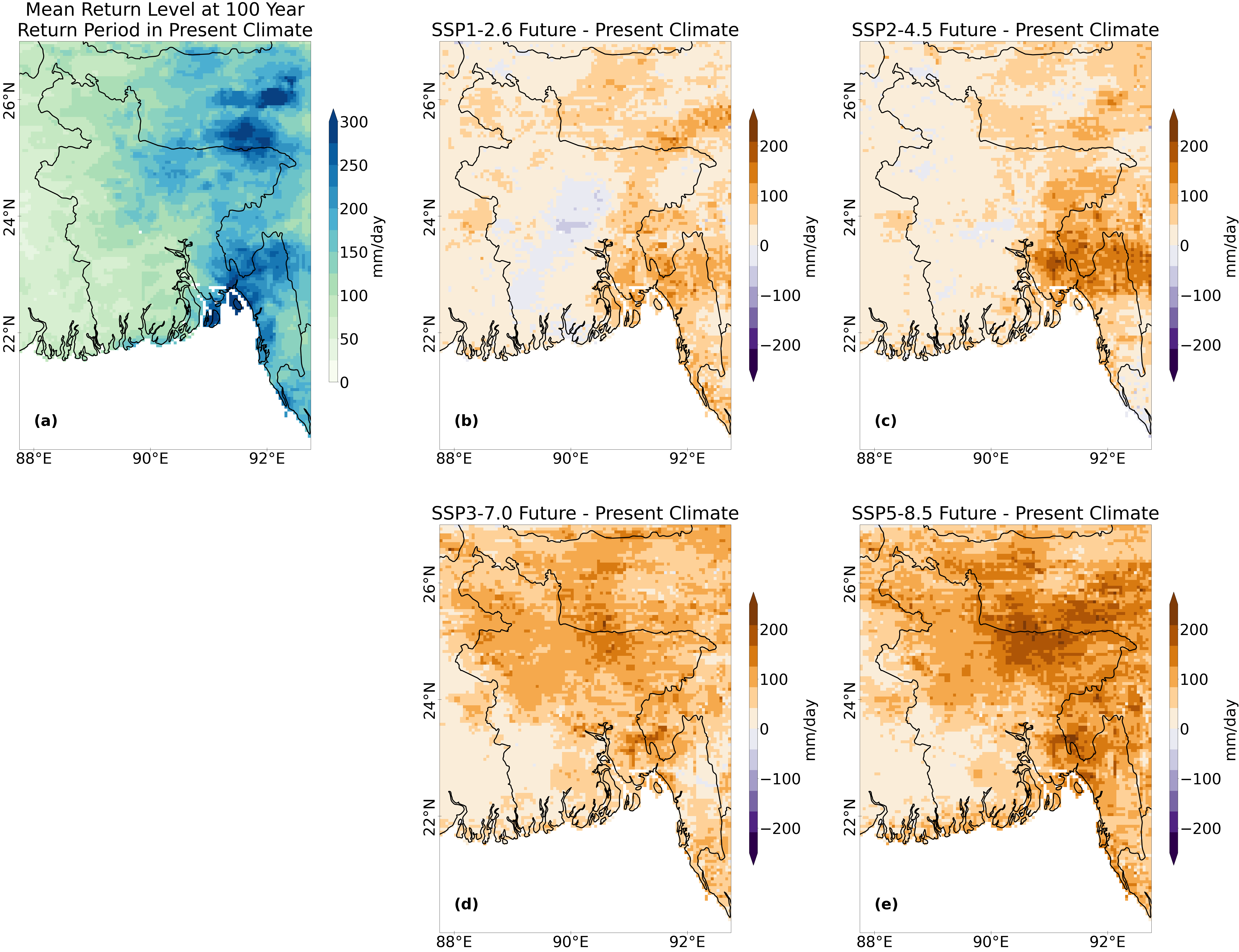}
\caption{Spatial distribution of extreme rainfall risk by the end of the century, captured by downscaled climate models. Panel (a) shows the mean return level at a 100-year return period for the present climate (1985-2014). Panel (b-e) displays the differences in the mean return level at a 100-year return period between the future (2081-2100) and present climates for four different scenarios: (b) SSP1-2.6, (c) SSP2-4.5, (d) SSP3-7.0, and (e)SSP5-8.5.}
\label{fig:riskfuturespatialend}
\end{figure*}

Figure~\ref{fig:riskfuturespatialmid} and Figure~\ref{fig:riskfuturespatialend} illustrate the spatial distribution of extreme rainfall risk captured by downscaled climate models, respectively, by the mid-century and end of the century. Comparing the mean return level for a 100-year return period between the present and future climates helps identify areas where the risk of extremes increases and the extent of this increase. The maximum rise in extreme risk will likely occur in northeast Bangladesh, a region already susceptible to extremes due to its surrounding topography, and the southeast coastal region, which is vulnerable to tropical cyclones. However, model projections and scenarios have substantial variability.

\section{Discussion}
\label{section:discussion}

Downscaling an ensemble of CMIP6 climate models shows that Bangladesh's risk of extreme events will increase significantly by the end of the century, especially for scenarios with heavy fossil fuel-driven development. This finding aligns with the global consensus on climate change. However, climate models do not have consensus on a regional scale~\cite{saha2019can}, and the results show substantial variability across models and warming scenarios. The spatial distribution of risk projects an increase in risk in the areas already heavily exposed to extremes.

A fundamental assumption of our method is that the downscaling function remains invariant to climate change, which allows the model to be trained on present-day data and applied to future scenarios. However, this assumption requires further validation. Training the model on low- and high-resolution simulations that cover current and future climates could address the issue. However, such simulations are not readily available due to computational cost. Additionally, investigating the risk of cascading extreme events, such as extreme rainfall accompanied by cyclones, heatwaves, or floods, is an important area for future research. Our downscaling framework extends to variables like wind, temperature, and inundation to provide a more integrated approach to future climate risk assessment. Furthermore, we are interested in the development of continuous scaling laws for model fields using geometry-coupled random field models~\cite{ravela2014spatial, ravela2015dynamic} and neural dynamics~\cite{trautner2020informative}.

\clearpage
\section*{Acknowledgment}
This research is part of the MIT Climate Grand Challenge Jameel Observatory CREWSNet and Weather and Climate Extremes projects. Schmidt Sciences, LLC and Liberty Mutual (029024-00020) provided support. The authors thank Dr. Jiangchao Qiu for the discussions.

\section*{Data Availability Statement}
All datasets used in this study are publicly available. CHIRPS global daily rainfall dataset is available from the Climate Hazards Center, University of California, Santa Barbara, USA (\url{https://data.chc.ucsb.edu/products/CHIRPS-2.0/}). the Copernicus Climate Change Service (C3S) Climate Data Store (CDS) (\url{https://cds.climate.copernicus.eu}) provides ERA5 and ERA5-land reanalysis data. The Earth System Grid Federation (ESGF) nodes (\url{https://aims2.llnl.gov}) provide HighResMIP data. The topographic elevation data obtained by Shuttle Radar Topography Mission (SRTM) at 90 m horizontal resolution from OpenTopography (\url{https://portal.opentopography.org/raster?opentopoID=OTSRTM.042013.4326.1}).

\bibliographystyle{unsrtnat}
\bibliography{references}
\end{document}